\documentclass[structuredabstract]{aa} 
\usepackage{graphicx}
\pdfoutput=1

\usepackage{txfonts}
\usepackage[utf8]{inputenc}
\usepackage{natbib}
\bibpunct{(}{)}{;}{a}{}{,} 

\def\aap{A\&A\ }

\def\apjl{ApJL\ }

\def\apjs{ApJS\ }

\begin{document}
   \title{Near-infrared integral-field spectra of the planet/brown dwarf companion AB Pic b\thanks{Based on service-mode observations (080.C-0590(A)) collected at the European Organisation for Astronomical Research in the Southern Hemisphere, Chile.}}

   \author{M.Bonnefoy
          \inst{1}
          \and
          G. Chauvin\inst{1}
         \and
          P. Rojo\inst{2}
         \and
          F. Allard\inst{3}
         \and
          A.-M. Lagrange\inst{1}
         \and
		 D. Homeier \inst{4}
         \and
          C. Dumas\inst{5}
         \and
          J.-L. Beuzit\inst{1}
         }
 
   \institute{Laboratoire d'Astrophysique, Observatoire de Grenoble, 414, Rue de la piscine, Saint-Martin d'H\`eres, France\\
              \email{mbonnefo@obs.ujf-grenoble.fr}
         \and
             Departamento de Astronomia, Universidad de Chile, Casilla 36-D, Santiago, Chile
         \and
             CRAL-ENS, 46, All\'ee d'Italie , 69364 Lyon Cedex 07
         \and
		     Institut für Astrophysik Göttingen, Georg-August-Universität, Friedrich-Hund-Platz 1, D-37077 Göttingen, Germany
         \and
             European Southern Observatory, Casilla 19001, Santiago 19, Chile
             }

   \date{Received 2009 June 12; accepted 2009 December 22}

\authorrunning{Bonnefoy et al.}
\titlerunning{NIR integral-field spectra of the planet/brown dwarf companion AB Pic b}

 
  \abstract
   {Chauvin et al. 2005 imaged a co-moving companion at a projected
     separation of $\sim$260 AU from the young star AB Pic
     A. Evolutionary models predictions based on $JHK$ photometry of AB
     Pic b suggested a mass of $\sim13-14~M_{\rm{Jup}}$, placing the object at
     the deuterium-burning boundary.}
   {We aim at determining the spectral type, the surface gravity and
     the effective temperature of AB Pic b. From the comparison of our
     absolute photometry to surface fluxes generated by atmospheric
     models, we also aim at deriving mass and radius estimates
     independent from evolutionary models predictions to test
     and refine them.}
   {We used the adaptive-optics-fed integral field spectrograph
 SINFONI to obtain high quality medium-resolution spectra of AB
 Pic b ($R_\lambda=1500-2000$) over the $1.1-2.5~\mu$m range. Our
 analysis relies on the comparison of our spectra to young
 standard templates and to the latest libraries of synthetic
 spectra developed by the Lyon's Group.}
  {AB Pic b is confirmed to be a young early-L dwarf companion. We
 derive a spectral type L0--L1 and find several features
 indicative of intermediate gravity atmosphere. A comparison to
    synthetic spectra yields $T_{\rm{eff}}=2000^{+100}_{-300}$~K and
    log$(g) = 4 \pm 0.5$ dex. The determination of the derived
 atmospheric parameters of AB Pic\,b is limited by a non-perfect
    match of current atmosphere spectra with our near-infrared
    observations of AB Pic\,b. The current treatment of dust settling
    and missing molecular opacity lines in the atmosphere models could be responsible. By
 combining the observed photometry, the surface fluxes from
    atmosphere models and the known distance of the system, we derive
    new mass, luminosity and radius estimates of AB Pic~b. They
 confirm independently the evolutionary model predictions. We finally review the current methods used to
 characterize planetary mass companions and discuss them in
 the perspective of future planet deep imaging surveys that will be
 confronted to the same limitations.}
	{}

   \maketitle
%

\section{Introduction}

Understanding how planets form and evolve and what physical processes
affect their atmospheric chemistry remain a major challenge of
exoplanetary science, since the first glimpse of planetary formation
revealed by the discovery of the $\beta$ Pictoris star debris disk
\citep{1984Sci...226.1421S}. Radial-velocity and transit searches
  have detected more than 300 exoplanets and, in some favorable cases,
  enabled to initiate the spectroscopic characterization of the
  irradiated atmospheres of transiting giant planets
  \citep{2004ApJ...604L..69V, 2009IAUS..253..231T,
    2009AIPC.1094..417B}. However, these techniques remain limited to
  the study of close-in planets with orbital radii typically smaller
  than 5 AU.

At wider orbits, the deep imaging technique is particularly well
suited using the space telescope (HST) or the combination of Adaptive
Optics (AO) system with very large ground-based telescopes (Palomar,
CFHT, Keck, Gemini, Subaru, VLT). The recent identification of young
($\le100$~Myr), nearby ($\le100$~pc) stars, members of comoving groups
\citep{1997Sci...277...67K, 2004ARA&A..42..685Z,
  2008hsf2.book..757T} has offered ideal targets to probe the
existence of substellar companions. Over the past years, several brown
dwarf companions have been successfully detected at relatively wide
($\ge40$~AU) orbits from their central stars
\citep{1999ApJ...512L..69L, 2000ApJ...541..390L,
  2003A&A...404..157C}. The planetary mass companion imaged at
$\sim41$ AU from the young nearby brown dwarf 2M1207
\citep{2004A&A...425L..29C} finally opened the way to new discoveries
on other classes of targets, members of distant open clusters (Itoh et
al. 2005, Neuh\"auser et al. 2005; Luhman et al. 2006; Lafreni\`ere et
al. 2008; Schmidt et al. 2008) and nearby intermediate-age
($0.1-1.0$~Gyr) stars (Metchev et al. 2006) and to the recent
breakthrough discoveries of HR\,8799\,bcd (Marois et al. 2008b),
Fomalhaut~b (Kalas et al. 2008) and the $\beta$ Pic\,b candidate (Lagrange et
al. 2008). In most cases, the companionship was generally confirmed
  using follow-up observations to show that the star and its companion
  share common proper motion, whereas the companion mass was always
  assumed from a comparison of the companion photometry with
  evolutionary model predictions at the system age and distance. In
  some cases, high quality spectra have enabled to derive the
  companion spectral type to confirm its relatively cool atmosphere
  with the identification of broad molecular absorptions.  The
  effective temperature $T_{\rm{eff}}$ and the surface gravity
  log$(g)$ could sometimes be derived using temperature- and
  gravity-sensitive features \citep{2003ApJ...593.1074G,
    2007ApJ...657..511A} by comparison to
  template spectra of field and young dwarfs \citep{Mclean2003, 2005ApJ...623.1115C, 2008MNRAS.383.1385L} or using atmosphere
  models. However, systematic and homogeneous photometric and
  spectroscopic characterization of young wide planetary mass
  companions are essential to further constrain interior and
  atmosphere models that could depend on their formation mechanisms
  \citep{2007ApJ...655..541M}.

\begin{table*}[t]
\begin{minipage}[ht]{\linewidth}
\caption{Observing log}             
\label{table:1}      
\centering          
\begin{tabular}{ l l l l l l l l l l  l l l}     
\hline\hline       
UT Date			& Target			& Grating 	& R$_{\lambda}$ 			&	Pre-optic 				& sec \textit{z}		& FWHM			 		& $\langle$EC$\rangle$ & 	$\langle\tau_{0}\rangle$	& DIT	 & NDIT & $t_{exp}$	& Note \\
				 	&			 			&			 	&		 		&	  (mas/pixel)			&							&   ('')					&   (\%)							&		(ms)		  						&  (s) 	 &			& 	(s)				&		   \\
\hline                    
05/12/2007 &	AB Pic b			&	   $J$   		& 	2000	&  25 $\times$ 12.5	&	1.225/1.221		&	  1.05/0.81		&	54.4							&	60.13								&	300	 &	 1		& 2700		&	   \\	
05/12/2007	&	AB Pic b			&	  $ J$   		& 	2000	&  25 $\times$ 12.5 	&	1.199/1.198		&	  0.72/0.83		&	54.6							&	71.73								&	300	 &	 1		& 	2700		&		   \\
05/12/2007	&	HIP039640	&	   $J$		& 	2000	&  25 $\times$ 12.5	&	1.202/1.200		&	  1.34/1.20		&	38.9							&	120.91 							&	40	 &	 1		&	80			& Tel STD	 \\
11/12/2007	&	AB Pic b			&	   $J$   		& 	2000	&  25 $\times$ 12.5	&	1.199/1.198		&	  0.82/0.81		&	51.5							&	66.16								&	300	 &	 1		&	2700		&			 \\
11/12/2007	&	HIP023230	&	   $J$   		& 	2000	&  25 $\times$ 12.5	&	1.214/1.218		&	  0.96/1.24		&	39.5							&	133.44 							&	60	 &	 1		&	120			&	Tel STD		 \\			
12/11/2007	&	AB Pic b			&	  $H+K$  	&	1500	&  25 $\times$ 12.5	&	1. 200/1.201		&	  1.67/1.74		&	14.9							&	11.15								&	300	 &	 1		&	2700		&					 \\
12/11/2007	&	HIP037963	&	  $H+K$  	& 	1500	&  25 $\times$ 12.5	&	1.211/1.210		&	  2.32/2.31		&	05.4							&	 6.11								&	20	 &	 3		&  120			&	Tel STD	 \\				
\hline                  
\end{tabular}
\end{minipage}
\end{table*}

In the course of a VLT/NACO deep coronographic imaging survey of
young, nearby stars \citep{2009Chauvin}, \cite{2005A&A...438L..25C} discovered a
faint commoving source at 5.5$~\!''$ ($\sim$260 AU) from the young star AB Pic\,A \citep[HIP30034, K2V, $V=9.16$,
d=45.5$_{-1.7}^{+1.8}$ pc,][]{1997A&A...323L..49P}. AB Pic A has been originally identified as
a member of the Tucana-Horologium association (Tuc-Hor) from its
distance to the Earth, the strength of the $\lambda$6708 Li line \citep[EW=260 $\pm$ 20 m\AA; confirmed by][]{2008ApJ...689.1127M}, its
filled-in H$\alpha$ absorption and saturated L/L$_{X}$ emission
\citep[hereafter S03]{2003ApJ...599..342S}. S03 also obtained galactic
space motions and found them to be similar to those of newly
identified Tuc-Hor members. At the age of the association \citep[$\sim30$~Myrs, see ][]{2000AJ....120.1410T, 2000A&A...361..581S,
  2000ApJ...535..959Z, 2001ApJ...559..388Z, 2001ASPC..244...43T,
  2007ApJ...662.1254S, 2008ApJ...689.1127M}, several evolutionary
models predict a companion mass of 13--14 $M_{\rm{Jup}}$ in agreement
with the L0--L3 spectral type derived from NACO $K$-band
spectroscopy. This places AB Pic b at the planetary mass boundary
($\sim$ 13.6 $M_{\rm{Jup}}$\footnote{\small~Following the definition
  of the \textit{International Astronomical Union.}}). Recently,
\citet{2007ApJS..169..105M} claimed that AB pic could be a member of
the $\beta$ Pictoris association ($\beta$ Pic) due to a lower
metallicity compared to Tuc-Hor members and a study of mutual
conjunctions with the $\rho$ Ophiucus and Upper Sco star forming
regions. The system would be therefore considerably younger. However,
this classification must be re-considered in the light of recent
\citet{2009arXiv0904.1221V} results assigning a solar metallicity to
AB Pic\,A (V. Makarov, Private Com.). Alternatively,
\citet{2008hsf2.book..757T} classified AB Pic A as a member of the 30
Myr-old Carina association from its enhanced L/L$_{X}$ emission and a
convergence kinematical method described in
\citet{2006A&A...460..695T}. In all cases, this membership revision
does not modify the age of AB Pic~A and b and the conclusion of
\citet{2005A&A...438L..29C}.

We present here high quality near-infrared spectra of the AB Pic b companion. They constitute the first results of a homogeneous survey to built an empirical library of carefully processed near-infrared spectra of young very-low mass companions. The observations of AB Pic b and the associated
data-reduction are presented in section~2. The analysis of our spectra
is presented in section~3. In section~4, we derive new mass, radius,
and luminosity estimations using various methods to confront their
respective limitations in the perspective of future deep imaging
search of giant planets.


\section{Observations and data reduction procedures}

	\subsection{Observations}

We used the SINFONI instrument \citep{2004Msngr.117...17B}, installed
on the Very Large Telescope UT4 (Yepun) to conduct a spectral analysis
of AB Pic b over the 1.1--2.45 $\mu$m range. The instrument benefits
from the high angular resolution provided by a modified version of the
Multi-Applications Curvature Adaptive Optic (AO) system MACAO
\citep{2003SPIE.4839..329B} and of the integral field spectroscopy
offered by SPIFFI \citep[SPectrograph for Infrared Faint Field Imaging, see][]{2003SPIE.4841.1548E}. AB Pic b was observed with the $J$
(1.1--1.4 $\mu$m) and $H+K$ (1.45--2.45 $\mu$m) gratings at resolving
powers of 2000 and 1500 respectively (see Table \ref{table:1}). The
AO-loop was closed on the bright AB Pic A (R=8.61). The instrument was
used with the 25 mas/pixel pre-optic, that limits the field of view
(FoV) to $0.8~\!'' \times 0.8~\!''$. Small dithering of the source (8
different positions) increased the FoV up to $1.1~\!''$ and allowed
filtering of residual bad pixels (see Part 2.2). At the end of each
observation, the telescope was nodded on the sky. Telluric standard
(STD) stars were observed with identical setups without dithering
right after AB Pic b.  Flat fields, darks, arc-lamp and distortion
calibration frames were obtained the days following the observations.

	\subsection{Data processing}

The whole dataset was reduced using custom \textit{IDL} routines and
the SINFONI data reduction pipeline version 1.9.8
\citep{2007astro.ph..1297M}. Routines were developed to correct raw
images from negatives rows created during the bias subtraction (see
\textit{The ESO data reduction cookbook}; version 1.0, ESO, 2007) and
to suppress the odd even effect affecting slitlet \#25. The pipeline
carried out cube reconstruction from bias and odd even effect
corrected images.  Hot and non-linear pixels were first flagged. The
distortion, the wavelength scale, and the slitlet positions were then
computed on the entire detector using arc-lamp frames. Slitlet
distances were measured with North-South scanning of the detector
illuminated with an optical fiber. Object-sky frame pairs were
subtracted, flat fielded and corrected from bad pixels and
distortions. Datacubes were finally reconstructed from clean science
images and were merged into a master cube. Filtering of sky line
residuals resulting from sky variation was first done with the
pipeline. A second iteration was performed on the slices of each
individual datacubes before merging then in the final master
cube. Finally, an additional routine was implemented to detect and
interpolate residual bad pixels.

AB Pic b datacubes did not suffer from the contamination of the nearby
AB pic A but were rather dominated by noise. Atmospheric refraction
produces a shift of the source with wavelength in the FoV (from 12.5
to $\sim70$~mas). The source was then re-centered slice per slice
using polynomial-fit and sub-pixelic shifts with bicubic-spline
magnification. The flux of the source was then integrated over an
aperture minimizing the noise without introducing differential flux
losses (radii of 187.5 to 512.5~mas). The STD datacubes were first
divided by a black body curve at the $T_{\rm{eff}}$ of the star and the
flux was collected over the entire FoV. The STD spectra were then
corrected from intrinsic features using a Legendre polynomial
interpolation over the surrounding continuum. AB Pic b spectra were
then divided by STD spectra, normalized and averaged to form a final
spectrum.


\section{Spectral analysis}

          \subsection{Empirical comparison}

   \begin{figure}[t]
   \centering
   \includegraphics[width=\columnwidth]{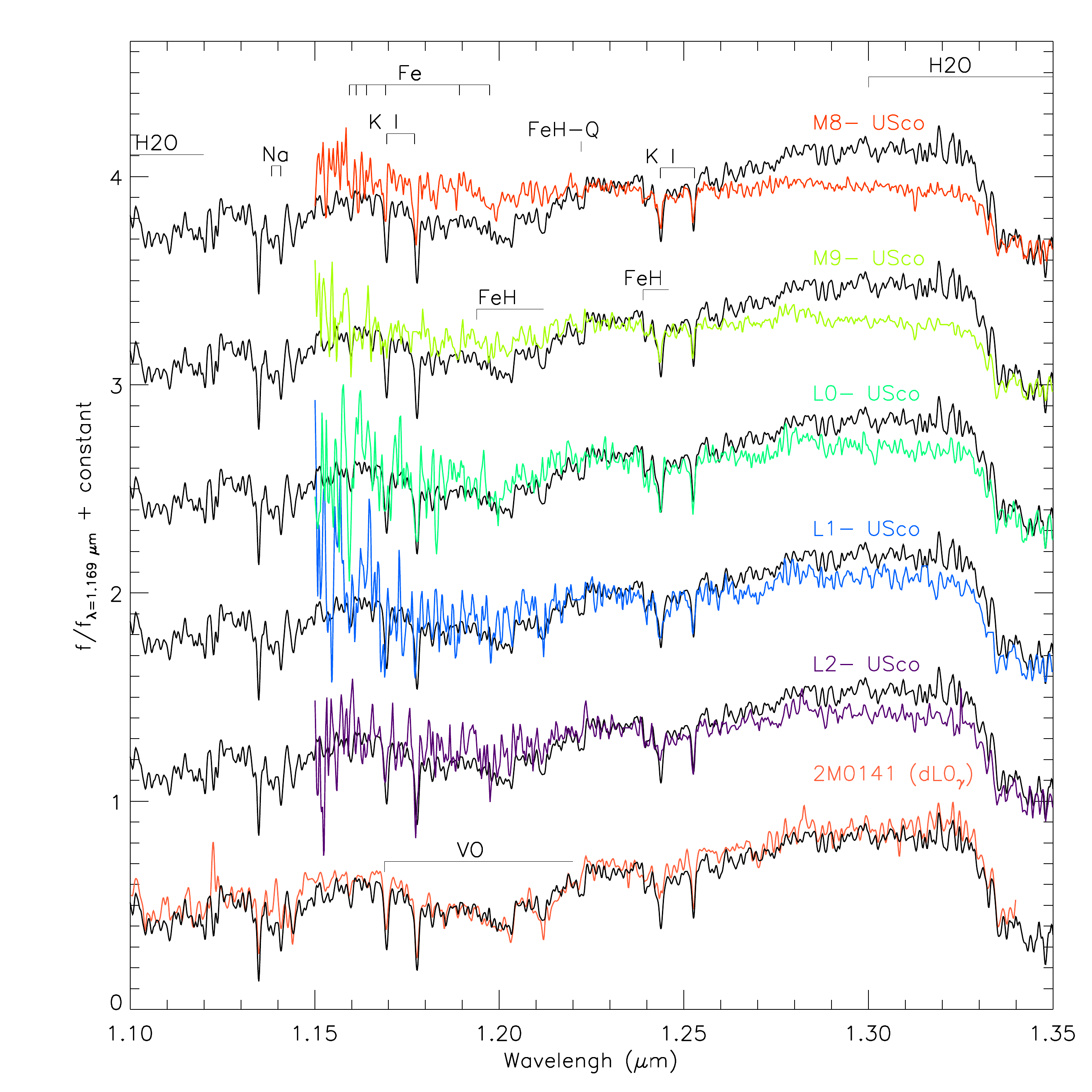}
\caption{$J$-band SINFONI spectrum (1.1--1.4 $\mu$m) of AB Pic b (black) compared to young ($\sim$ 5 Myrs) Upper-Sco brown-dwarf spectra (color) over the 1.15--1.35 $\mu$m range and to the SINFONI spectrum (salmon-pink) of the young isolated object 2M0141 (classified dL0$_{\gamma}$). AB Pic b and 2M0141 spectra were convolved with a Gaussian kernel to match the templates resolution (R=1400). $\chi^{2}$ are minimized for young L1 and L2 dwarfs. The weakened K I lines at 1.169, 1.177, 2.243 and 2.253 $\mu$m, the enhanced VO band from 1.17 to 1.22 $\mu$m and the reduced FeH band around 1.2 $\mu$m confirms the youth of the object.}
         \label{Jband}
   \end{figure}

   \begin{figure}[t]
   \centering
   \includegraphics[width=\columnwidth]{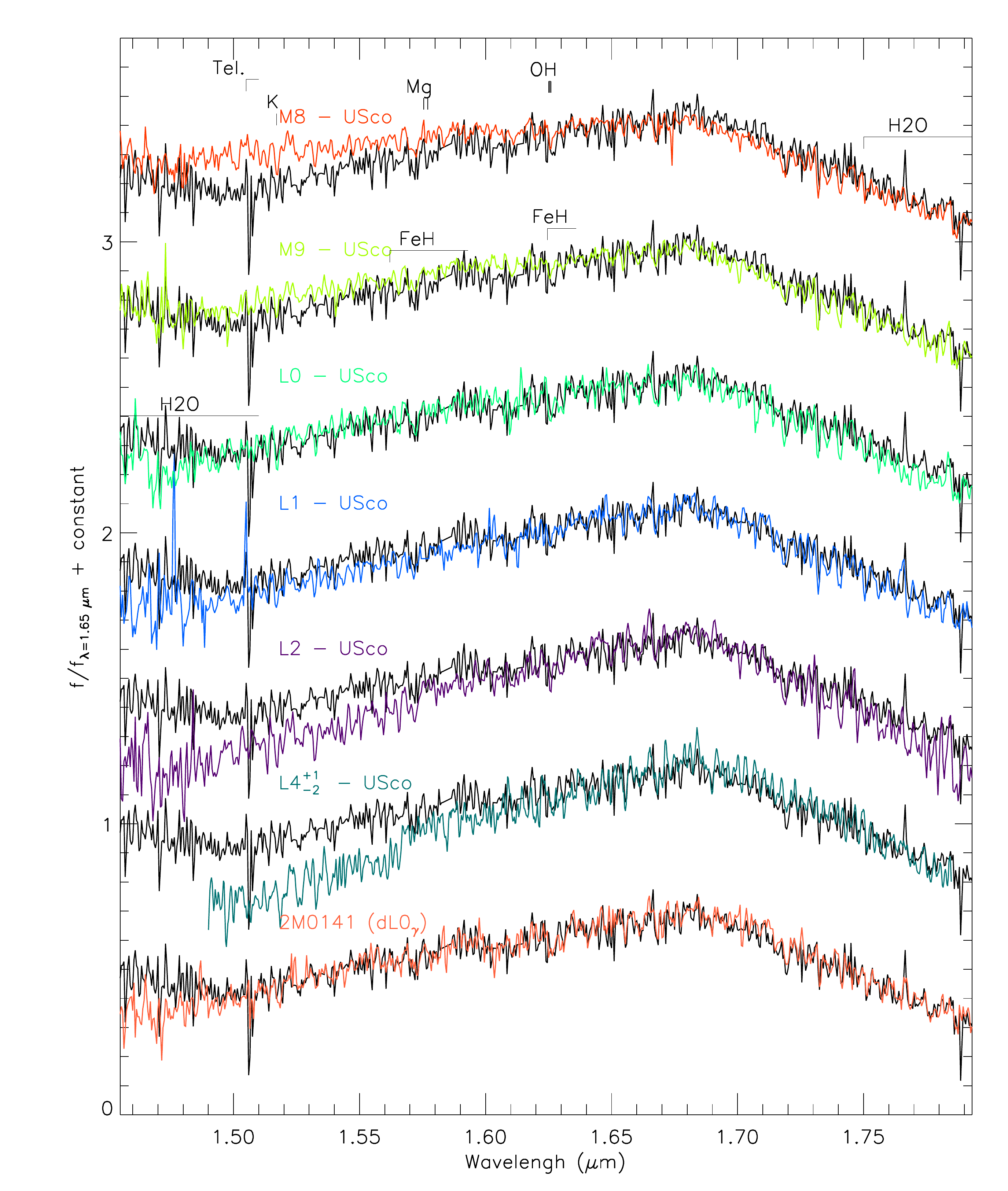}
\caption{$H$-band SINFONI spectrum (1.45--1.8 $\mu$m) of AB Pic b (black) compared to young ($\sim$ 5 Myrs) Upper-Sco brown-dwarf spectra (color) over the 1.5--1.8 $\mu$m range and to the SINFONI spectrum (salmon-pink) of the young isolated object 2M0141 (classified dL0$_{\gamma}$). $\chi^{2}$ are minimized for a young L0 dwarf. The triangular-shaped profile of our spectrum results from reduced collision induced absorption of H$_{2}$ observed in young objects. The K I line at 1.517 $\mu$m and the FeH absorption around 1.63 $\mu$m observed in L0 field dwarf spectra are not present in our spectrum.}
         \label{Hband}
   \end{figure}

\begin{figure}[t]
   \centering
   \includegraphics[width=\columnwidth]{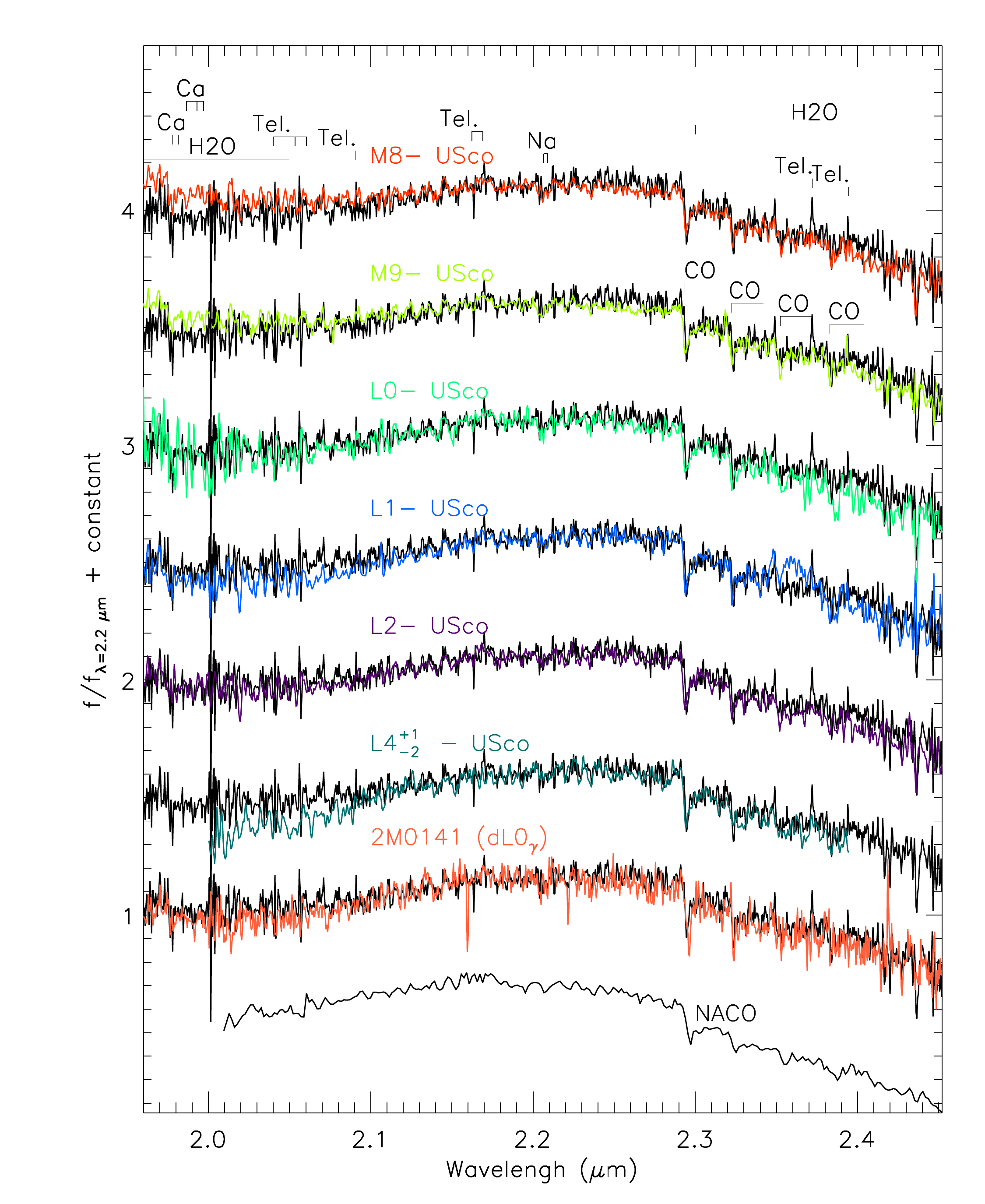}
\caption{$K$-band SINFONI spectrum of AB Pic b (black) compared to young ($\sim$ 5 Myrs) Upper-Sco brown-dwarf spectra (color) and to the young isolated object 2M0141 (classified dL0$_{\gamma}$). $\chi^{2}$ are minimized for young L0 and L2 dwarfs. The shape is not well reproduced by old field dwarfs. The low-resolution ($R_{\lambda}$=550) NACO spectrum is overplotted at the bottom for comparison.}
         \label{Kband}
   \end{figure}
\begin{figure}[t]
   \centering
   \includegraphics[width=\columnwidth]{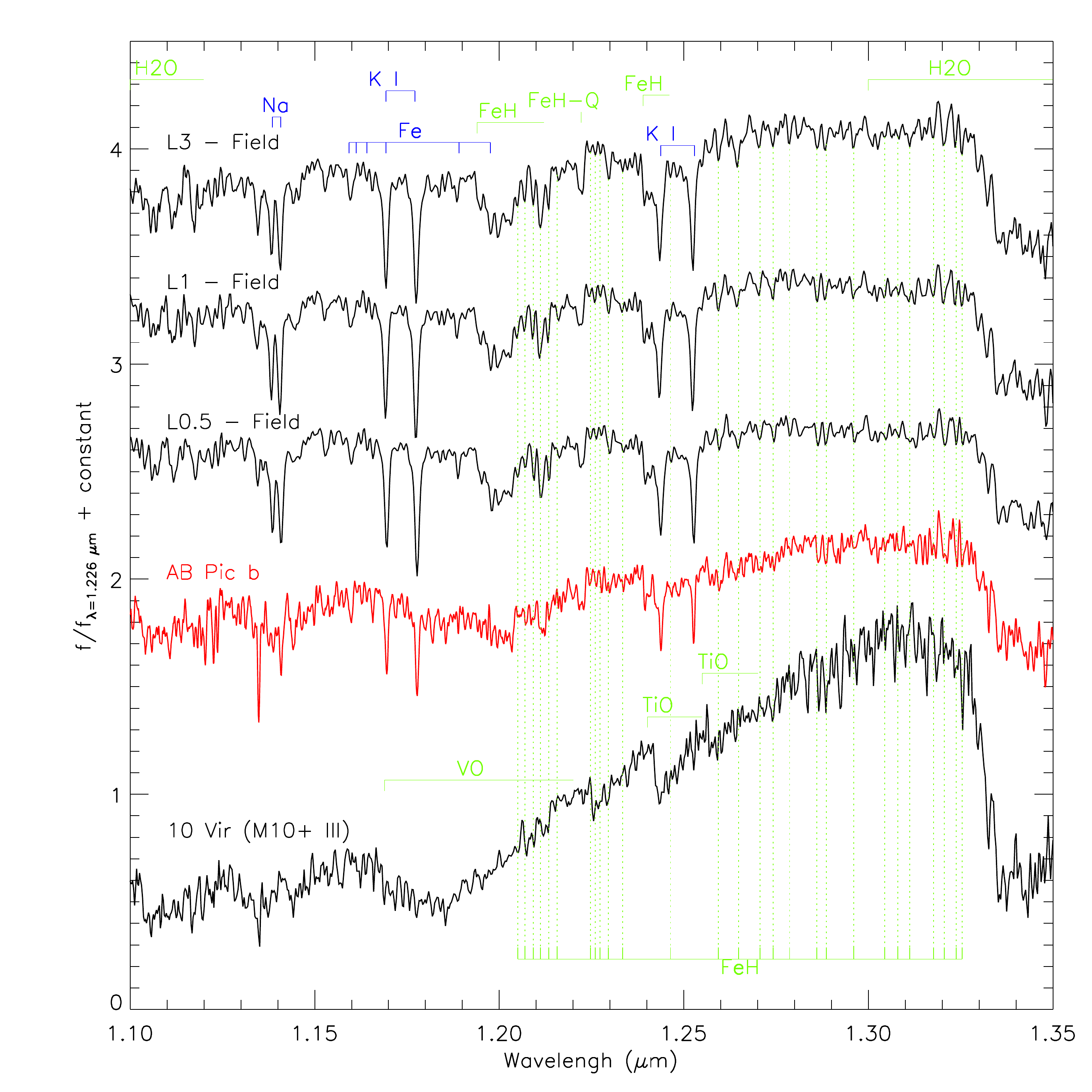}
\caption{$J$-band SINFONI spectrum of AB Pic b (red) compared to  the spectrum of a late-type M giant (10 Vir) and to field dwarf spectra \citep{2005ApJ...623.1115C} over the 1.1--1.35 $\mu$m range. Atomic features are plotted in blue and molecular absorptions in green. Our spectrum shows features intermediate between the dwarfs and the giant. It is best reproduced by the spectrum of 2MASSJ1439+1929 (L1).}
         \label{Jbandfield}
   \end{figure}

The $J$, $H$ and $K$-band normalized spectra of AB Pic b are presented
in Fig.~\ref{Jband}, \ref{Hband}, \ref{Kband}. The signal to noise ratio ranges from 40 in the H band to 50 in the J and K bands. They are compared to
young M8-L2 brown dwarfs spectra of Upper Sco ($\sim$ 5 Myrs)
\citep[hereafter L08]{2008MNRAS.383.1385L} at identical resolution. In
addition, $H$ and $K$-band spectra are compared to the spectrum of the
L$4^{+1}_{-2}$ companion candidate to 1RXS J160929.1-210524, member of
Upper Sco \citep{2008ApJ...689L.153L}. In agreement with our
expectations, the AB Pic b spectra are typical of young late-M/early-L
dwarfs. The triangular shape of the $H$-band
\citep{2001MNRAS.326..695L, 2004ApJ...617..565L}, the bumpy $K$-band,
the reduced strength of alkali lines (Na I at 1.138 $\mu$m, K I
doublets at 1.169/1.177 and 1.243/1.253 $\mu$m and the K I line at
1.517 $\mu$m) and of FeH absorptions over the $J$ and $H$ bands
\citep{2001ApJ...559..424W, Cushing} are all resulting from the
intermediate surface-gravity of the companion. This is well
illustrated in Fig.~\ref{Jbandfield}, where the $J$-band spectrum of
AB Pic~b shows intermediate-gravity KI doublets compared to low and
high-gravity spectra of the 10~Vir giant star and field dwarfs.

To assign a spectral type, we used two different approaches.  We first compared our spectra to those of young M8--L4 brown dwarfs \citep[from L08
and ][]{2008ApJ...689L.153L} classified in the near-infrared. A $\chi^{2}$ minimization was obtained
for young L1 and L2 dwarfs in the $J$ band, L0 dwarf in the $H$ band
and, L0 and L2 dwarfs in the $K$ band. L2 and L4 were excluded from the visual
comparison of the pseudo-continuum shape in the H-band. We also used spectral indexes to measure
the depth of water absorptions at 1.34 $\mu$m \citep[H$_{2}$OA and
H$_{2}$O--1; see][]{Mclean2003, 2004ApJ...610.1045S},
1.5 $\mu$m \citep{2007ApJ...657..511A}, and 2.04 $\mu$m \citep[H$_{2}$O--2,
see][]{2004ApJ...610.1045S}. These absorbtions are only slightly
age-insensitive. These indexes confirmed that AB Pic~b has a near-infrared spectral
type intermediate between a L0 and L1 (Table~\ref{table:2}). The K1
and K2 indexes measuring the strength of the H$_{2}$O band from 2.0 to
2.14 $\mu$m and of the H$_{2}$ absorption around 2.2 $\mu$m
respectively \citep[hereafter T99]{1999AJ....117.1010T} finally
discriminated AB Pic b from old field dwarfs (see Fig.~4 of T99;
computed for young BD and field dwarfs). 

However, near-infrared spectral types are not necessarily consistent with those infered at optical wavelenghs where a classification scheme homogeneous for field and young brown dwarfs exists. We noticed that the normalised spectra of the L08 sample tend to be bluer than that of AB Pic b in the J band. AB Pic b normalized $H+K$ band spectrum appear in addition overluminous in the $K$ band (or underluminous in the $H$-band). The only exeption are UScoJ163919-253409 and UScoJ160918-222923 (classified as L1 by L08) for which a good simultaneous fit in the J and H+K bands is achieved. This discrepency could be related to peculiar dust properties, differences in dust opacities, binarity or even extinction by a disk \citep{2007ApJ...666.1219L}. However, \cite{2009ApJ...696.1589H} revealed  that the classification of L dwarfs given in L08 could be strongly revised using optical spectra. We then ultimately decided to compare our spectra to those of the young L0 dwarf 2MASS J01415823-4633574 \citep[hereafter K06]{2006ApJ...639.1120K}. This dwarf (hereafter 2M0141) has been classified in the optical following the scheme of \citet{2005ARA&A..43..195K}. The excellent match of our spectrum with that of 2M0141 in the J, and H+K bands (see Fig. \ref{Jband}, \ref{Hband}, and \ref{Kband}) leads us to assign a final spectral type L0--L1 to AB Pic\,b.

 \begin{table}[t]
\caption{Spectral indexes and associated spectral types for AB Pic b. We reported indices designed to classify dwarfs: H$_{2}$OA \citep{Mclean2003} and the equivalent H$_{2}$O--1 \citep{2004ApJ...610.1045S}, H$_{2}$O--2 \citep{2004ApJ...610.1045S}, H$_{2}$O \citep{2007ApJ...657..511A}. K1 and K2 indexes \citep{1999AJ....117.1010T} show that AB Pic b is a young L dwarf (L$_{pec}$).}             
\label{table:2}      
\centering                          
\begin{tabular}{l l l l l l l}     
\hline\hline        
\textbf{Indice} & H$_{2}$OA & H$_{2}$O--1 & H$_{2}$O--2 & H$_{2}$O & K1 & K2\\         
\hline
\textbf{Value} & 0.57 & 0.61 & 0.89 & 1.14 & 0.1 & 0.04 \\
\textbf{Sp. type} & L$2\pm2$ & L$2\pm2$ & L$0\pm2$ & M$9\pm2$& L$_{pec}$ & L$_{pec}$ \\
\hline                                   
\end{tabular}
\end{table}

A $T_{\rm{eff}}$-spectral type conversion scales have not been established for
young early-L dwarfs yet. However, the \cite{2003ApJ...593.1093L}
scale valid for young M dwarfs ($\sim$ 2 Myrs) gives an upper limit
fot the effective temperature of $2400$~K. L08 give in addition an
estimation of 2000 K $\lesssim$ $T_{\rm{eff}} \lesssim 2300$ K for L0
to L1 young brown-dwarfs of the Upper Sco association.

               \subsection{Comparison to atmosphere grids}

\begin{figure}[t]
   \centering
   \includegraphics[width=\columnwidth]{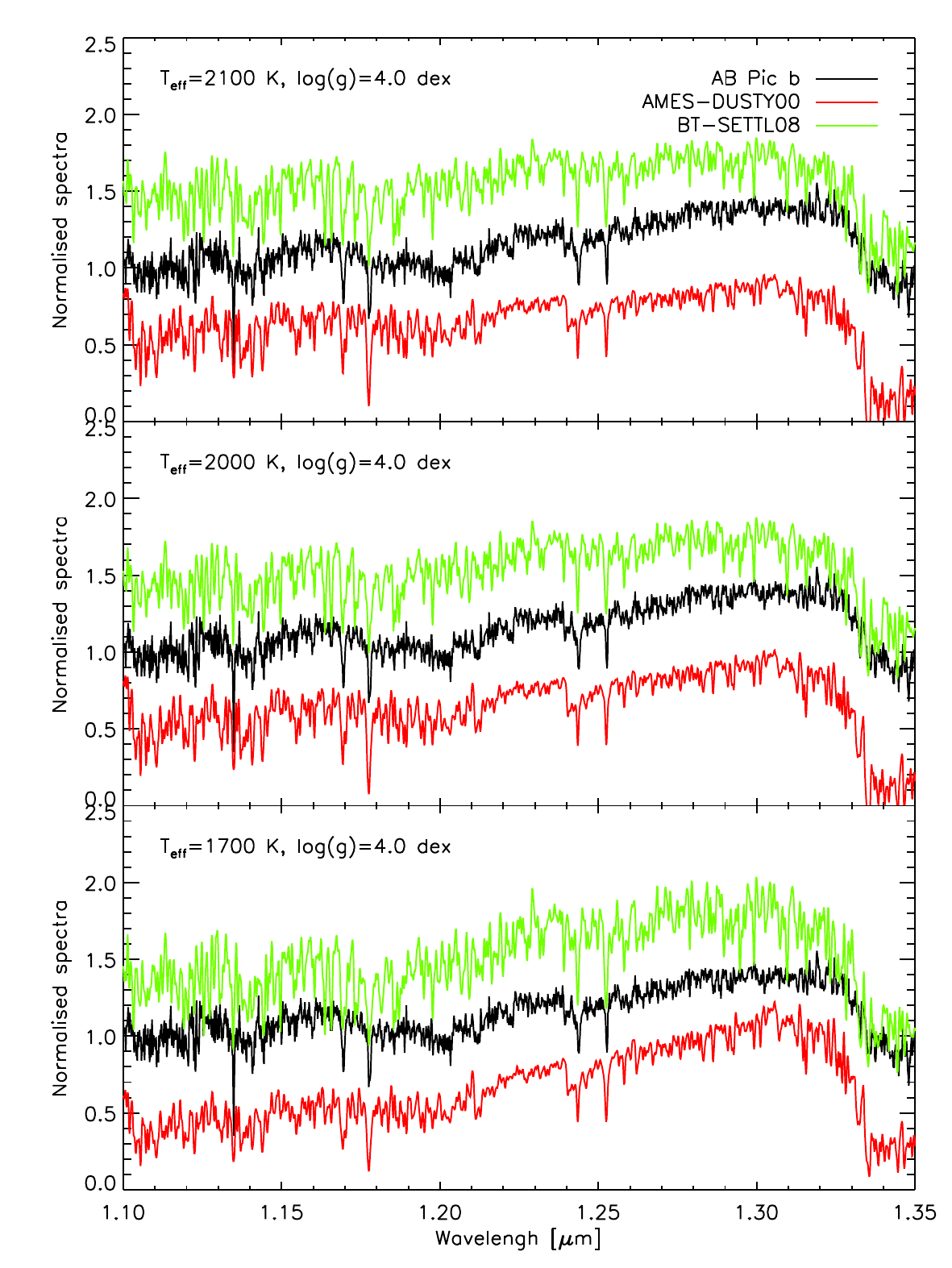}
	\caption{Comparison of the AB Pic b J band spectrum to synthetic spectra of the AMES-Dusty00 and SETTL08 libraries at $log(g)$=4.0 dex, [M/H]=0 dex, and $T_{eff}$=1700 K, 2000 K and 2100 K.}
         \label{Jbandcomp}
   \end{figure}

\begin{figure}[t]
   \centering
   \includegraphics[width=\columnwidth]{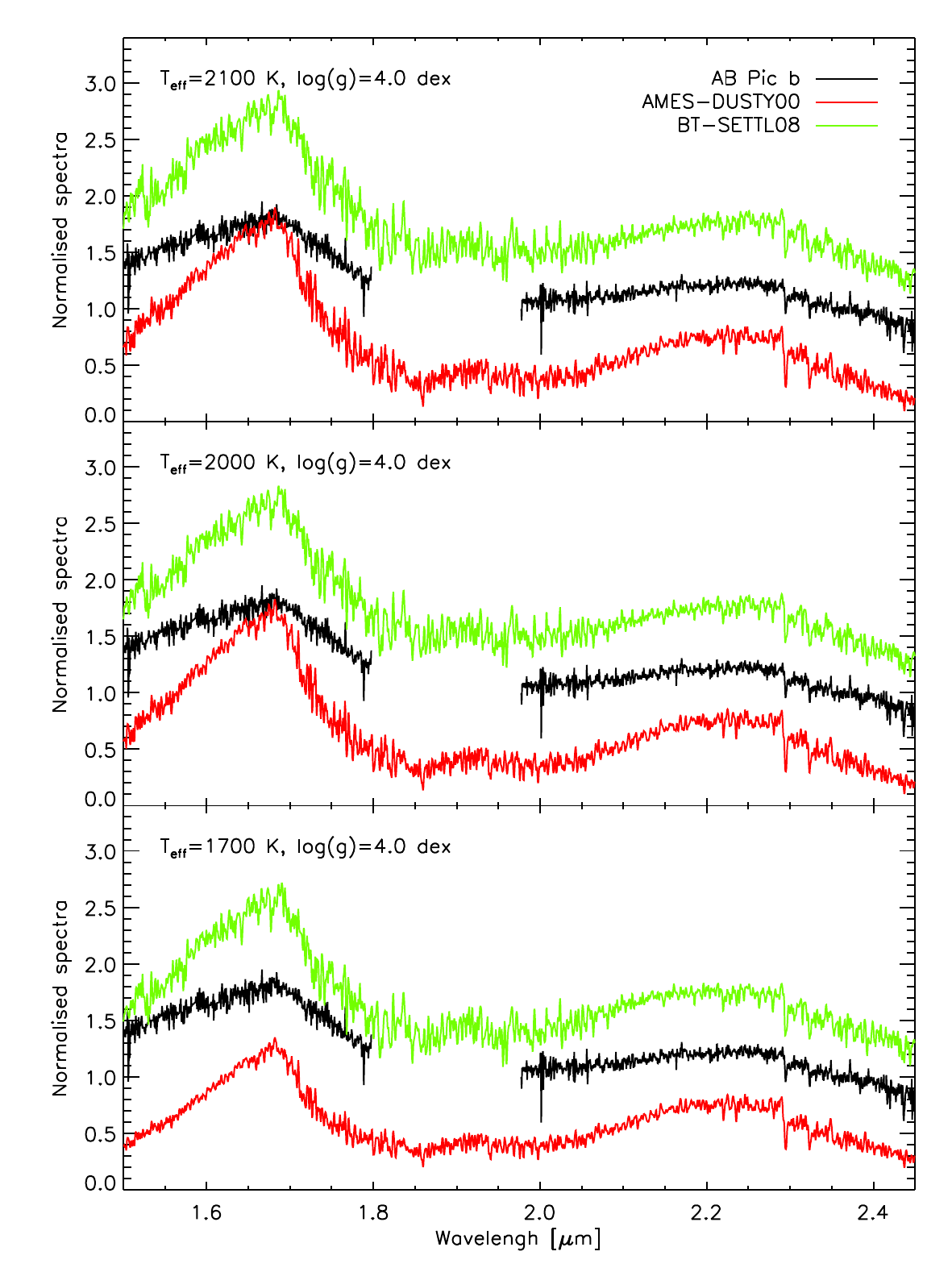}
	\caption{Comparison of the AB Pic b H+K band spectrum to synthetic spectra of the AMES-Dusty00 and SETTL08 libraries at $log(g)$=4.0, [M/H]=0 dex, and $T_{eff}$=1700 K, 2000 K and 2100 K.}
         \label{HKbandcomp}
   \end{figure}

We used the AMES--Dusty00 library of synthetic spectra\footnote{The
  spectra can be generated on demand on
  \textit{http://phoenix.ens-lyon.fr/simulator/index.faces}}
\citep{2001ApJ...556..357A} that incorporates formation of dust in the
atmosphere for $T_{\rm{eff}} \lesssim 2600$~K. Spectra were convolved
by a Gaussian kernel to match the resolution of SINFONI, interpolated
on the AB Pic b wavelength grid and normalized at 1.226 $\mu$m, 1.56
$\mu$m and 2.2 $\mu$m in the $J$, $H$ and $K$ band. H$_{2}$O
absorptions from 1.32 to 1.60 $\mu$m and from 1.75 to 2.20 $\mu$m are
known to be overestimated in the models \citep{2001ApJ...548..908L,
  2001MNRAS.326..695L, 2006ApJ...639.1120K}. We then compared the AB
Pic b spectral continuum to spectral models over the 1.1--1.34~$\mu$m
zone using a classical weighted least-square method. The analysis led
to $T_{\rm{eff}}=2000\pm100$~K and log$(g)=4.0\pm 0.5$ dex. The fit
was still visually acceptable for $T_{\rm{eff}}$ down to 1700 K and up
to 2100 K (see Figs. \ref{Jbandcomp} and \ref{HKbandcomp}). Comparatively, K06 found similar atmospheric parameters for
2M0141.

The $K$-band was not properly fitted at $T_{\rm{eff}}=2000$ K and
higher effective temperatures ($T_{\rm{eff}}=2500$ K) could better
match the observed depth of CO overtones at $\lambda\geqslant 2.3
\mu$m and the shape of the pseudo-continuum. The case of 2M0141 is
likely to be similar (see the lower panel of Fig~9 of K06). Finally,
the $K$-band of AB Pic b was better reproduced at
$T_{\rm{eff}}=2000$~K and log$(g)=4.0$~dex using the most recent SETTL08 library. This library incorporates updated bank of
molecular opacities (BT), a more realistic mixing length parameter
($\alpha$=2) and settling and replenishment of the dust in the
photosphere. 

Finally, the $H$-band is marginally reproduced by AMES--Dusty00 and SETTL08 models. These non-reproducibilities are strenghtened in SETTL08 spectra by a bad representation of the thermal structure of the photosphere. This effect is also responsible for the overestimation of several narrow absorptions in  the J band (see Fig. \ref{Jbandcomp}).

\begin{figure}[t]
   \centering
   \includegraphics[width=\columnwidth]{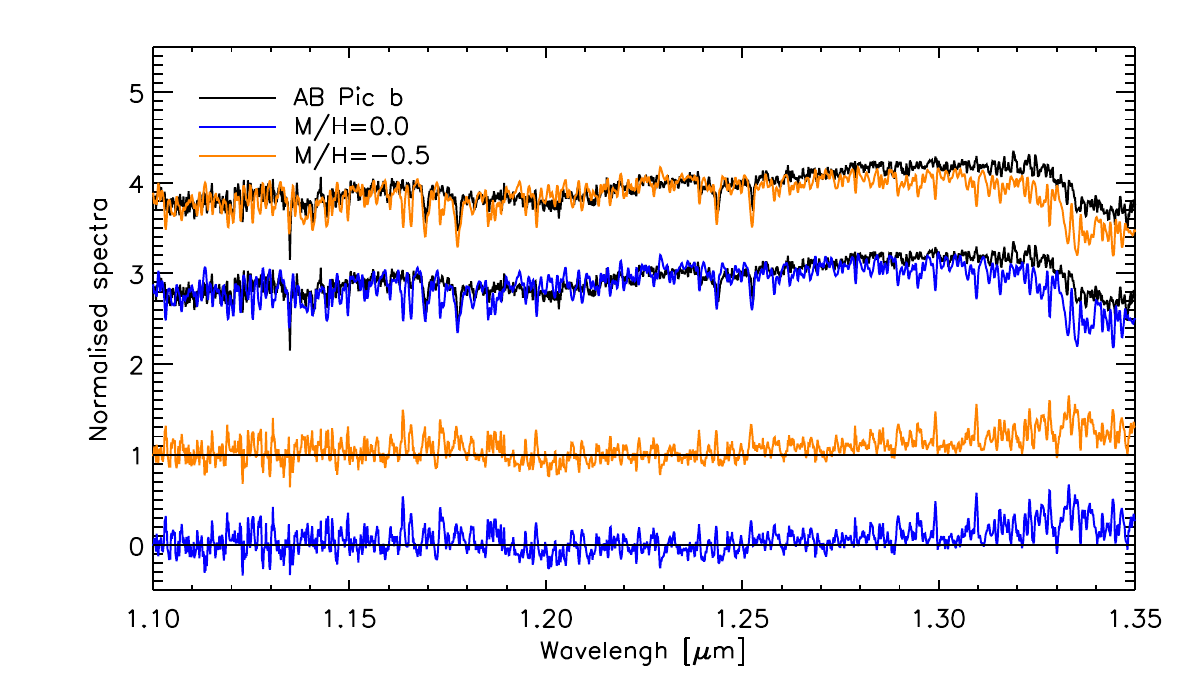}
	\caption{Comparison of the AB Pic b J band spectrum to normalised synthetic spectra (along with residuals) of the SETTL08 library at $log(g)$=4.0, $T_{eff}$=2000 K and  --0.5 $<$ [M/H] $<$ 0.0. The pseudo continuum is not strongly affected by the metallicity. However, we can notice that the depth of the K I doublets around 1.169 and 1.243 $\mu$m increases for subsollar abundances.}
         \label{Jbandcompmet}
   \end{figure}

\begin{figure}[t]
   \centering
   \includegraphics[width=\columnwidth]{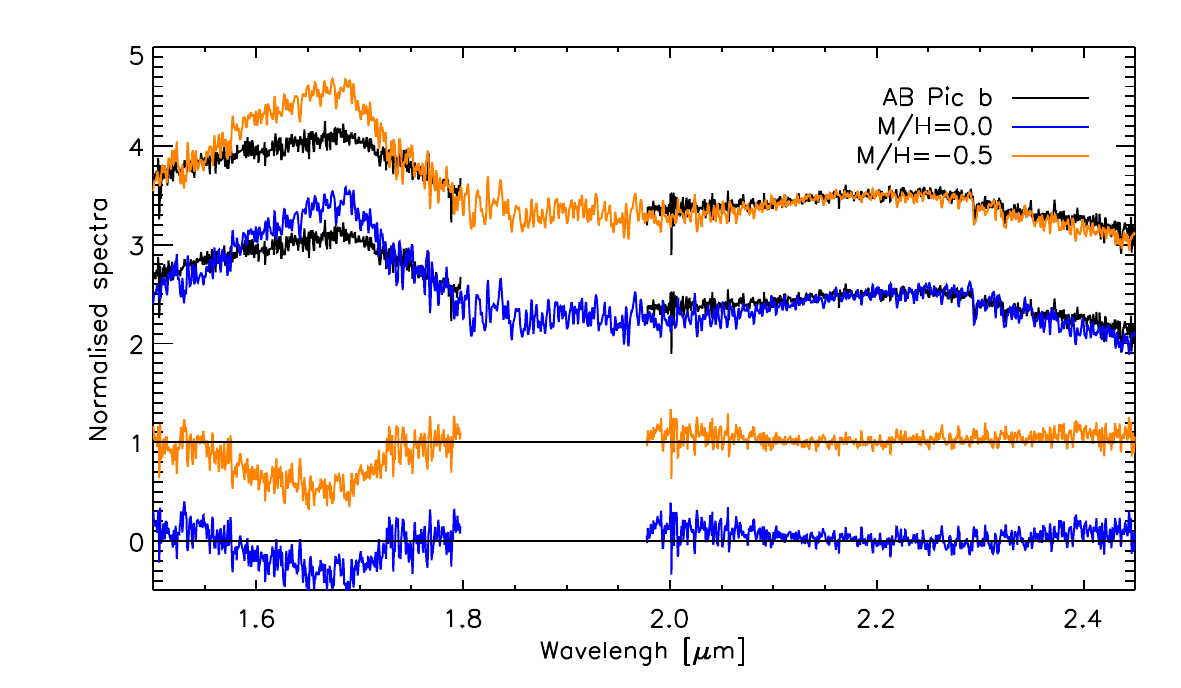}
	\caption{Same as Fig.~\ref{Jbandcompmet} but for the H+K band. Interestingly, it can be seen that spectra at [M/H]=--0.5 better reproduce the depth of CO overtones found in AB Pic b spectra.}
         \label{HKbandcompmet}
   \end{figure}

\citet{2004A&A...418..989N} reported AB Pic A as a metal poor star
with Fe/H$=-0.64 \pm0.12$~dex. However, this estimation relies on a
relation based on color indices that could be biased for very young
objects. More recently, \citet[hereafter V09]{2009arXiv0904.1221V}
measured spectroscopically a Fe/H$=+0.07$ dex. At a distance of $\sim$ 260 AU, one probable formation mechanism for AB Pic b is via a
binary-like process, i.e with initial abundances similar to those of
AB pic A. We then compared our spectrum to SETTL08 models with Fe/H
between --0.5 and 0.0 and $T_{\rm{eff}}=2000^{+100}_{-300}$ K to
test their influence on our fits. The models indicate that sub-solar
abundances do not change the $J$-band pseudo-continuum but increase
the depth of the K I doublets at 1.169/1.177 $\mu$m and 1.243/1.253
$\mu$m (see Fig~\ref{Jbandcompmet}). The $K$-band shows weaker CO overtones and reduced $H_{2}O$ absorptions (see Fig~\ref{HKbandcompmet}). This effect does not affect our $T_{\rm{eff}}$ estimation but could lead to slightly lower log$(g)$ estimates.  Further abundances analysis of AB Pic~b should add new constraints. 

In conclusion, both AMES--Dusty00 and SETTL08 libraries yield
$T_{\rm{eff}}=2000^{+100}_{-300}$ K and log$(g)=4.0 \pm 0.5$~dex for AB
Pic b, for solar and sub-solar (Fe/H=--0.5) metallicities. In comparison, \cite{2007ApJ...657.1064M} (hereafter M07) fit near-infrared colors and absolute photometry of AB Pic b with SETTL05 and AMES-DUSTY00 models. They found a good mach for  $T_{\rm{eff}}$=1700--1800 K and log(g)=4.25. Similarily, the SETTL08 grid reproduces the observed colors for $T_{\rm{eff}}$=1600--1700 K and log(g)=4.0--4.5. However, the lack of reproducibilities of synthetic spectra in the H and K band revealed here shows that near-infrared colors and absolute fluxes  should be used with care to infer the properties of young and cool objects.   

\section{Radius, mass, and luminosity of AB Pic~b}

                    \subsection{Evolutionary models predictions}

Masses of young substellar companions are mostly derived from
evolutionary models predictions \citep{1997MmSAI..68..807D,
  1997ApJ...491..856B, 1998A&A...337..403B, 2000ApJ...542..464C,
  2003A&A...402..701B, 2008ApJ...689.1327S}. State of the art models
are based on the combination of interior models and atmospheric models
necessary to link up the luminosities in observed bands to the mass,
radius, effective temperatures and surface gravities at different ages
and metallicities (tracks). From the de-reddened magnitudes and colors
in the 2MASS passbands, the AMES--Dusty00 evolutionary models
(\citet{2000ApJ...542..464C}; using AMES--Dusty00 atmospheric models)
predict for AB Pic~b a mass of 10~$M_{\rm{Jup}}$ $\leqslant$ M
$\leqslant$ 14~$M_{\rm{Jup}}$. Using $T_{\rm{eff}}$ derived from our
spectral analysis leads to a similar prediction of 11~$M_{\rm{Jup}}$
$\leqslant$ M $\leqslant$ 14~$M_{\rm{Jup}}$. Scarce direct mass
measurements of young brown dwarfs and very low mass stars seem to
reveal that models does not achieve a good simultaneous prediction of
$T_{\rm{eff}}$, luminosities, radii and masses of young low mass
objects \citep[see ][]{2007prpl.conf..411M}. Tracks remain to be
calibrated at young ages and very low masses down to the planetary
mass regime where the formation mechanisms could actually play a key
role \citep{2007ApJ...655..541M}.  One can therefore try to use
alternative methods to estimate the mass of substellar companions.

                    \subsection{Alternative estimations}

Based on the empirical relations between BC$_{K}$ and spectral types,
the luminosity of the object can be independently determined from
evolutionary tracks. In the case of a L0--L1 spectral type, we
derived a bolometric correction BC$_{K}$=$3.24^{+0.18}_{-0.19}$~mag
from the relations of \citet{2004AJ....127.3516G} valid for field
dwarfs. Luminosity, radius and mass can then be deduced based on
$T_{\rm{eff}}$ and log$(g)$ derived from our spectral analysis (see
Table~\ref{table:3}). The estimated mass is in agreement with
evolutionary model predictions but with much larger uncertainties.
However, the assumption of a similar BC$_{K}$-SpT relation between
young and field dwarfs has never been established and could therefore
add systematic errors that are difficult to quantify.

An alternative approach has been introduced by \citet[hereafter
  M04]{2004ApJ...609..854M} to estimate masses, radii and luminosities
of isolated substellar objects of the Upper Sco association. The
method is to use the surface flux provided by atmospheric models at
the distance of the system. In the specific case of AB Pic b,
considering a $A_{V}=0.27\pm0.02$ \citep[][see]{2009ApJ...694.1085V}, the absolute magnitude in $K$-band can
therefore be combined to the surface flux of the AMES--Dusty00 spectra at
$T_{\rm{eff}}=2000$ K and log$(g)=4.0$ (computed in the 2MASS filters)
to determine the companion radius.  From the radius and the surface
gravity, we can deduce the mass, and considering in addition
$T_{\rm{eff}}$, we derive the luminosity. All results are reported in
Table \ref{table:3}. The same approach can be done using $J$-band flux
leading to slightly different values explained by a non-simultaneous
reproducibility of the $J$ and $K_s$-band surface flux at
$T_{\rm{eff}}=2000$ K.  This effect has already been noticed in M04 for
derivations of the radii, luminosities and masses from $I_{C}$ and $J$
magnitudes. In both cases, estimated masses of AB Pic~b are again in
agreement with evolutionary model predictions within
uncertainties. The errors on these parameters results from propagated
errors of the atmospheric parameters of our spectral analysis (that
make the surface flux varying) and of the absolute photometry. Similar
results are derived from the SETTL08 surface fluxes.

\begin{table}[t]
\caption{Updated AB Pic b properties. They were first re-estimated from evolutionary models (line 1), based on empirical BC$_{K}$ of field dwarfs (line 2), finally, using surface fluxes in the $J$ and $K$-band predicted by AMES--Dusty00 atmospheric models (lines 3 and 4).}             
\label{table:3}      
\centering          
\small                
\begin{tabular}{llllll}     
\hline\hline  \noalign{\smallskip}      
\textbf{Method} &     $T_{\rm{eff}}$       &    log$(g)$     &       log(L/L$_{\odot}$)                &     Radius                 &  Mass              \\
                &     (K)                  &                 &               &     (R$_{Jup}$)            &  (M$_{Jup}$)       \\                  
\hline\noalign{\smallskip}
Evol.           &      $1750_{-100}^{+100}$           & $4.2\pm0.2$       &       $-3.6\pm0.2$        &        1.5--1.6            &   10--14             \\
\noalign{\smallskip}BC$_K$          &      $2000_{-300}^{+100}$ &  4.0 $\pm$ 0.5 &  $-3.7\pm 0.2$   & $1.22^{+0.70}_{-0.25}$     &   1--45              \\
\noalign{\smallskip}$F_{\rm{surf-K}}$&     $2000_{-300}^{+100}$ &  4.0 $\pm$ 0.5 &  $-3.72_{-0.20}^{+0.15}$  & $1.13_{-0.11}^{+0.38}$     &   2--24              \\
\noalign{\smallskip}$F_{\rm{surf-J}}$&     $2000_{-300}^{+100}$ &  4.0 $\pm$ 0.5 &  $-4.00_{-0.16}^{+0.33}$  & $0.81_{-0.2}^{+0.83}$     &   1--21              \\
\noalign{\smallskip}  \hline\noalign{\smallskip}                                   
\end{tabular}
\end{table}

Both methods lead to consistent masses, luminosities, surface
gravities, and $T_{\rm{eff}}$ with evolutionary model predictions, but  do not allow to actually test and refine them. Our
conclusions are very sensitive to the values of $T_{\rm{eff}}$ and
log$(g)$ inferred from our spectral analysis and therefore adequate
atmosphere spectra modeling. The strong overprediction of H$_{2}$O
absorptions that dominate the shape of near-infrared spectrum of L
dwarfs, even for the most recent atmospheric models, and the
non-simultaneous reproducibility of the $J$, $H$ and $K$ bands both
indicate that the comparison of near-infrared spectra of planetary
mass companions to synthetic spectra is not straight forward. It will
therefore remain a difficult task for future direct imaging detection
and characterization that will rely on lower resolution spectra in
near-infrared and similar libraries of synthetic spectra. For
planetary mass companions at very wide orbit (no expected
contamination from the primary), fitting the spectral energy
distribution over a broader spectral range and directly measuring the
companion luminosity with combined thermal $L-$ or $M-$band photometry
or spectroscopy would certainly lead to a more robust estimation of
mass and radius as recently obtained by \cite{2008ApJ...682.1256L}
in the case of HN Peg~B.


\section{Conclusions}

We obtained high quality 1.1--2.5 $\mu$m medium-resolution spectra of
the young very low mass companion AB Pic b which evolutionary models
place at the planet/brown-dwarf boundary. Near-infrared spectra of young M8--L4 dwarfs were compared to the spectrum of AB Pic b using least-square fitting and spectral indexes. They confirmed the youth of the object and allowed to refine the spectral type estimation of \cite{2005A&A...438L..29C} to L0--L1. Our spectral classification was also confirmed by the excellent mach of our spectrum to that of the young L0 field dwarf 2MASS J01415823-4633574 classified in the visible. A comparison to synthetic spectra enabled us to derive $T_{\rm{eff}}=2000^{+100}_{-300}$ K and log$(g)=4.0\pm0.5$ compatible with --0.5 $\lesssim$ Fe/H $\lesssim$ 0. Finally, we used bolometric correction valid for field L dwarfs and atmospheric models to estimate the mass, the luminosity and the radii of the companion independently from evolutionary models predictions. The lack of bolometric corrections for young L dwarfs and the large uncertainties related to the atmospheric parameters determination do not currently allow to refine the mass of AB Pic b, necessary to state on its status and to test evolutionary models predictions.

Our study points out the difficulties to infer young L dwarfs companion properties from spectral analysis alone. Alternative methods to evolutionary models used here rely on uncelebrated relations and atmospheric models. They remain to be tested on a large wavelength range at various gravities (age), effective temperatures and metallicities. These loopholes will strongly limit the characterization of gaseous planets detected with the upcoming planet-finders Gemini/GPI \citep{2006SPIE.6272E..18M} and VLT/SPHERE \citep{2006Msngr.125...29B,2008SPIE.7014E..41B}.  In that perspective, the XSHOOTER instrument at VLT will enable to acquire simultaneous spectra over 0.3--2.5~$\mu$m to robustly constrain $T_{\rm{eff}}$, log$(g)$, metallicity as well as luminosity of isolated and wide companions members of young nearby associations and star forming regions down to the planetary mass regime. The comparison of the observed luminosity to predictions of evolutionary models will also provide a crucial constraint of evolutionary models at young ages and planetary masses, and could also clarify the role of formation mechanisms in this mass range.

\begin{acknowledgements}

 We are very grateful to the anonymous referee for its constructive review that greatly improved our initial manuscript.
We thank the ESO Paranal staff for performing the service mode
 observations. We also acknowledge partial financial support from the
 \textit{Agence National de la Recherche} and the \textit{Programmes
   Nationaux de Planétologie et de Physique Stellaire} (PNP \& PNPS),
 in France. We are grateful to Andreas Seifahrt, David Lafrenière and
 Nicolas Lodieu for providing their spectra. Finally, this work would have not
 been possible without the NIRSPEC
 (\textit{http://www.astro.ucla.edu/}$\sim$\textit{mclean/BDSSarchive/}),
 IRTF
 (\textit{http://irtfweb.ifa.hawaii.edu/}$\sim$\textit{spex/IRTF}$\_Spectral\_Library/$)
 and SpecX (\textit{http://www.browndwarfs.org/spexprism}) libraries
 maintained by Ian S. McLean, Michael C. Cushing, John T. Rayner, and
 Adam Burgasser.

\end{acknowledgements}

\bibliographystyle{aa}
\bibliography{Bonnefoy_letter2009_ABPicb}

\begin{thebibliography}{66}
\expandafter\ifx\csname natexlab\endcsname\relax\def\natexlab#1{#1}\fi

\bibitem[{{Allard} {et~al.}(2001){Allard}, {Hauschildt}, {Alexander},
  {Tamanai}, \& {Schweitzer}}]{2001ApJ...556..357A}
{Allard}, F., {Hauschildt}, P.~H., {Alexander}, D.~R., {Tamanai}, A., \&
  {Schweitzer}, A. 2001, \apj, 556, 357

\bibitem[{{Allers} {et~al.}(2007){Allers}, {Jaffe}, {Luhman}, {Liu}, {Wilson},
  {Skrutskie}, {Nelson}, {Peterson}, {Smith}, \&
  {Cushing}}]{2007ApJ...657..511A}
{Allers}, K.~N., {Jaffe}, D.~T., {Luhman}, K.~L., {et~al.} 2007, \apj, 657, 511

\bibitem[{{Baraffe} {et~al.}(1998){Baraffe}, {Chabrier}, {Allard}, \&
  {Hauschildt}}]{1998A&A...337..403B}
{Baraffe}, I., {Chabrier}, G., {Allard}, F., \& {Hauschildt}, P.~H. 1998, \aap,
  337, 403

\bibitem[{{Baraffe} {et~al.}(2003){Baraffe}, {Chabrier}, {Barman}, {Allard}, \&
  {Hauschildt}}]{2003A&A...402..701B}
{Baraffe}, I., {Chabrier}, G., {Barman}, T.~S., {Allard}, F., \& {Hauschildt},
  P.~H. 2003, \aap, 402, 701

\bibitem[{{Barnes} {et~al.}(2009){Barnes}, {Barman}, {Jones}, {Leigh},
  {Cameron}, {Prato}, \& {Barber}}]{2009AIPC.1094..417B}
{Barnes}, J.~R., {Barman}, T.~S., {Jones}, H.~R.~A., {et~al.} 2009, in American
  Institute of Physics Conference Series, Vol. 1094, American Institute of
  Physics Conference Series, ed. E.~{Stempels}, 417--420

\bibitem[{{Beuzit} {et~al.}(2006){Beuzit}, {Feldt}, {Dohlen}, {Mouillet},
  {Puget}, {Antichi}, {Baruffolo}, {Baudoz}, {Berton}, {Boccaletti},
  {Carbillet}, {Charton}, {Claudi}, {Downing}, {Feautrier}, {Fedrigo}, {Fusco},
  {Gratton}, {Hubin}, {Kasper}, {Langlois}, {Moutou}, {Mugnier}, {Pragt},
  {Rabou}, {Saisse}, {Schmid}, {Stadler}, {Turrato}, {Udry}, {Waters}, \&
  {Wildi}}]{2006Msngr.125...29B}
{Beuzit}, J.-L., {Feldt}, M., {Dohlen}, K., {et~al.} 2006, The Messenger, 125,
  29

\bibitem[{{Beuzit} {et~al.}(2008){Beuzit}, {Feldt}, {Dohlen}, {Mouillet},
  {Puget}, {Wildi}, {Abe}, {Antichi}, {Baruffolo}, {Baudoz}, {Boccaletti},
  {Carbillet}, {Charton}, {Claudi}, {Downing}, {Fabron}, {Feautrier},
  {Fedrigo}, {Fusco}, {Gach}, {Gratton}, {Henning}, {Hubin}, {Joos}, {Kasper},
  {Langlois}, {Lenzen}, {Moutou}, {Pavlov}, {Petit}, {Pragt}, {Rabou}, {Rigal},
  {Roelfsema}, {Rousset}, {Saisse}, {Schmid}, {Stadler}, {Thalmann}, {Turatto},
  {Udry}, {Vakili}, \& {Waters}}]{2008SPIE.7014E..41B}
{Beuzit}, J.-L., {Feldt}, M., {Dohlen}, K., {et~al.} 2008, in Presented at the
  Society of Photo-Optical Instrumentation Engineers (SPIE) Conference, Vol.
  7014, Society of Photo-Optical Instrumentation Engineers (SPIE) Conference
  Series

\bibitem[{{Bonnet} {et~al.}(2004){Bonnet}, {Abuter}, {Baker}, {Bornemann},
  {Brown}, {Castillo}, {Conzelmann}, {Damster}, {Davies}, {Delabre},
  {Donaldson}, {Dumas}, {Eisenhauer}, {Elswijk}, {Fedrigo}, {Finger},
  {Gemperlein}, {Genzel}, {Gilbert}, {Gillet}, {Goldbrunner}, {Horrobin}, {Ter
  Horst}, {Huber}, {Hubin}, {Iserlohe}, {Kaufer}, {Kissler-Patig}, {Kragt},
  {Kroes}, {Lehnert}, {Lieb}, {Liske}, {Lizon}, {Lutz}, {Modigliani}, {Monnet},
  {Nesvadba}, {Patig}, {Pragt}, {Reunanen}, {R{\"o}hrle}, {Rossi}, {Schmutzer},
  {Schoenmaker}, {Schreiber}, {Stroebele}, {Szeifert}, {Tacconi}, {Tecza},
  {Thatte}, {Tordo}, {van der Werf}, \& {Weisz}}]{2004Msngr.117...17B}
{Bonnet}, H., {Abuter}, R., {Baker}, A., {et~al.} 2004, The Messenger, 117, 17

\bibitem[{{Bonnet} {et~al.}(2003){Bonnet}, {Str{\"o}bele}, {Biancat-Marchet},
  {Brynnel}, {Conzelmann}, {Delabre}, {Donaldson}, {Farinato}, {Fedrigo},
  {Hubin}, {Kasper}, \& {Kissler-Patig}}]{2003SPIE.4839..329B}
{Bonnet}, H., {Str{\"o}bele}, S., {Biancat-Marchet}, F., {et~al.} 2003, in
  Presented at the Society of Photo-Optical Instrumentation Engineers (SPIE)
  Conference, Vol. 4839, Adaptive Optical System Technologies II. Edited by
  Wizinowich, Peter L.; Bonaccini, Domenico. Proceedings of the SPIE, Volume
  4839, pp. 329-343 (2003)., ed. P.~L. {Wizinowich} \& D.~{Bonaccini}, 329--343

\bibitem[{{Burrows} {et~al.}(1997){Burrows}, {Marley}, {Hubbard}, {Lunine},
  {Guillot}, {Saumon}, {Freedman}, {Sudarsky}, \&
  {Sharp}}]{1997ApJ...491..856B}
{Burrows}, A., {Marley}, M., {Hubbard}, W.~B., {et~al.} 1997, \apj, 491, 856

\bibitem[{{Chabrier} {et~al.}(2000){Chabrier}, {Baraffe}, {Allard}, \&
  {Hauschildt}}]{2000ApJ...542..464C}
{Chabrier}, G., {Baraffe}, I., {Allard}, F., \& {Hauschildt}, P. 2000, \apj,
  542, 464

\bibitem[{{Chauvin} {et~al.}(2009){Chauvin}, {Lagrange}, {Bonavita},
  {Zuckerman}, {Dumas}, {Bessell}, J.-L., M., S., J., {Lowrance}, {Mouillet},
  \& {Song}}]{2009Chauvin}
{Chauvin}, G., {Lagrange}, A.-M., {Bonavita}, M.-A., {et~al.} 2009, \aap,
  submitted

\bibitem[{{Chauvin} {et~al.}(2004){Chauvin}, {Lagrange}, {Dumas}, {Zuckerman},
  {Mouillet}, {Song}, {Beuzit}, \& {Lowrance}}]{2004A&A...425L..29C}
{Chauvin}, G., {Lagrange}, A.-M., {Dumas}, C., {et~al.} 2004, \aap, 425, L29

\bibitem[{{Chauvin} {et~al.}(2005{\natexlab{a}}){Chauvin}, {Lagrange}, {Dumas},
  {Zuckerman}, {Mouillet}, {Song}, {Beuzit}, \&
  {Lowrance}}]{2005A&A...438L..25C}
{Chauvin}, G., {Lagrange}, A.-M., {Dumas}, C., {et~al.} 2005{\natexlab{a}},
  \aap, 438, L25

\bibitem[{{Chauvin} {et~al.}(2005{\natexlab{b}}){Chauvin}, {Lagrange},
  {Zuckerman}, {Dumas}, {Mouillet}, {Song}, {Beuzit}, {Lowrance}, \&
  {Bessell}}]{2005A&A...438L..29C}
{Chauvin}, G., {Lagrange}, A.-M., {Zuckerman}, B., {et~al.} 2005{\natexlab{b}},
  \aap, 438, L29

\bibitem[{{Chauvin} {et~al.}(2003){Chauvin}, {Thomson}, {Dumas}, {Beuzit},
  {Lowrance}, {Fusco}, {Lagrange}, {Zuckerman}, \&
  {Mouillet}}]{2003A&A...404..157C}
{Chauvin}, G., {Thomson}, M., {Dumas}, C., {et~al.} 2003, \aap, 404, 157

\bibitem[{{Cushing} {et~al.}(2003){Cushing}, {Rayner}, {Davis}, \&
  {Vacca}}]{Cushing}
{Cushing}, M.~C., {Rayner}, J.~T., {Davis}, S.~P., \& {Vacca}, W.~D. 2003,
  \apj, 582, 1066

\bibitem[{{Cushing} {et~al.}(2005){Cushing}, {Rayner}, \&
  {Vacca}}]{2005ApJ...623.1115C}
{Cushing}, M.~C., {Rayner}, J.~T., \& {Vacca}, W.~D. 2005, \apj, 623, 1115

\bibitem[{{D'Antona} \& {Mazzitelli}(1997)}]{1997MmSAI..68..807D}
{D'Antona}, F. \& {Mazzitelli}, I. 1997, Memorie della Societa Astronomica
  Italiana, 68, 807

\bibitem[{{Eisenhauer} {et~al.}(2003){Eisenhauer}, {Abuter}, {Bickert},
  {Biancat-Marchet}, {Bonnet}, {Brynnel}, {Conzelmann}, {Delabre}, {Donaldson},
  {Farinato}, {Fedrigo}, {Genzel}, {Hubin}, {Iserlohe}, {Kasper},
  {Kissler-Patig}, {Monnet}, {Roehrle}, {Schreiber}, {Stroebele}, {Tecza},
  {Thatte}, \& {Weisz}}]{2003SPIE.4841.1548E}
{Eisenhauer}, F., {Abuter}, R., {Bickert}, K., {et~al.} 2003, in Society of
  Photo-Optical Instrumentation Engineers (SPIE) Conference Series, Vol. 4841,
  Society of Photo-Optical Instrumentation Engineers (SPIE) Conference Series,
  ed. M.~{Iye} \& A.~F.~M. {Moorwood}, 1548--1561

\bibitem[{{Golimowski} {et~al.}(2004){Golimowski}, {Leggett}, {Marley}, {Fan},
  {Geballe}, {Knapp}, {Vrba}, {Henden}, {Luginbuhl}, {Guetter}, {Munn},
  {Canzian}, {Zheng}, {Tsvetanov}, {Chiu}, {Glazebrook}, {Hoversten},
  {Schneider}, \& {Brinkmann}}]{2004AJ....127.3516G}
{Golimowski}, D.~A., {Leggett}, S.~K., {Marley}, M.~S., {et~al.} 2004, \aj,
  127, 3516

\bibitem[{{Gorlova} {et~al.}(2003){Gorlova}, {Meyer}, {Rieke}, \&
  {Liebert}}]{2003ApJ...593.1074G}
{Gorlova}, N.~I., {Meyer}, M.~R., {Rieke}, G.~H., \& {Liebert}, J. 2003, \apj,
  593, 1074

\bibitem[{{Herczeg} {et~al.}(2009){Herczeg}, {Cruz}, \&
  {Hillenbrand}}]{2009ApJ...696.1589H}
{Herczeg}, G.~J., {Cruz}, K.~L., \& {Hillenbrand}, L.~A. 2009, \apj, 696, 1589

\bibitem[{{Kastner} {et~al.}(1997){Kastner}, {Zuckerman}, {Weintraub}, \&
  {Forveille}}]{1997Sci...277...67K}
{Kastner}, J.~H., {Zuckerman}, B., {Weintraub}, D.~A., \& {Forveille}, T. 1997,
  Science, 277, 67

\bibitem[{{Kirkpatrick}(2005)}]{2005ARA&A..43..195K}
{Kirkpatrick}, J.~D. 2005, \araa, 43, 195

\bibitem[{{Kirkpatrick} {et~al.}(2006){Kirkpatrick}, {Barman}, {Burgasser},
  {McGovern}, {McLean}, {Tinney}, \& {Lowrance}}]{2006ApJ...639.1120K}
{Kirkpatrick}, J.~D., {Barman}, T.~S., {Burgasser}, A.~J., {et~al.} 2006, \apj,
  639, 1120

\bibitem[{{Lafreni{\`e}re} {et~al.}(2008){Lafreni{\`e}re}, {Jayawardhana}, \&
  {van Kerkwijk}}]{2008ApJ...689L.153L}
{Lafreni{\`e}re}, D., {Jayawardhana}, R., \& {van Kerkwijk}, M.~H. 2008, \apjl,
  689, L153

\bibitem[{{Leggett} {et~al.}(2001){Leggett}, {Allard}, {Geballe}, {Hauschildt},
  \& {Schweitzer}}]{2001ApJ...548..908L}
{Leggett}, S.~K., {Allard}, F., {Geballe}, T.~R., {Hauschildt}, P.~H., \&
  {Schweitzer}, A. 2001, \apj, 548, 908

\bibitem[{{Leggett} {et~al.}(2008){Leggett}, {Saumon}, {Albert}, {Cushing},
  {Liu}, {Luhman}, {Marley}, {Kirkpatrick}, {Roellig}, \&
  {Allers}}]{2008ApJ...682.1256L}
{Leggett}, S.~K., {Saumon}, D., {Albert}, L., {et~al.} 2008, \apj, 682, 1256

\bibitem[{{Lodieu} {et~al.}(2008){Lodieu}, {Hambly}, {Jameson}, \&
  {Hodgkin}}]{2008MNRAS.383.1385L}
{Lodieu}, N., {Hambly}, N.~C., {Jameson}, R.~F., \& {Hodgkin}, S.~T. 2008,
  \mnras, 383, 1385

\bibitem[{{Lowrance} {et~al.}(1999){Lowrance}, {McCarthy}, {Becklin},
  {Zuckerman}, {Schneider}, {Webb}, {Hines}, {Kirkpatrick}, {Koerner}, {Low},
  {Meier}, {Rieke}, {Smith}, {Terrile}, \& {Thompson}}]{1999ApJ...512L..69L}
{Lowrance}, P.~J., {McCarthy}, C., {Becklin}, E.~E., {et~al.} 1999, \apjl, 512,
  L69

\bibitem[{{Lowrance} {et~al.}(2000){Lowrance}, {Schneider}, {Kirkpatrick},
  {Becklin}, {Weinberger}, {Zuckerman}, {Plait}, {Malmuth}, {Heap}, {Schultz},
  {Smith}, {Terrile}, \& {Hines}}]{2000ApJ...541..390L}
{Lowrance}, P.~J., {Schneider}, G., {Kirkpatrick}, J.~D., {et~al.} 2000, \apj,
  541, 390

\bibitem[{{Lucas} {et~al.}(2001){Lucas}, {Roche}, {Allard}, \&
  {Hauschildt}}]{2001MNRAS.326..695L}
{Lucas}, P.~W., {Roche}, P.~F., {Allard}, F., \& {Hauschildt}, P.~H. 2001,
  \mnras, 326, 695

\bibitem[{{Luhman} {et~al.}(2007){Luhman}, {Adame}, {D'Alessio}, {Calvet},
  {McLeod}, {Bohac}, {Forrest}, {Hartmann}, {Sargent}, \&
  {Watson}}]{2007ApJ...666.1219L}
{Luhman}, K.~L., {Adame}, L., {D'Alessio}, P., {et~al.} 2007, \apj, 666, 1219

\bibitem[{{Luhman} {et~al.}(2004){Luhman}, {Peterson}, \&
  {Megeath}}]{2004ApJ...617..565L}
{Luhman}, K.~L., {Peterson}, D.~E., \& {Megeath}, S.~T. 2004, \apj, 617, 565

\bibitem[{{Luhman} {et~al.}(2003){Luhman}, {Stauffer}, {Muench}, {Rieke},
  {Lada}, {Bouvier}, \& {Lada}}]{2003ApJ...593.1093L}
{Luhman}, K.~L., {Stauffer}, J.~R., {Muench}, A.~A., {et~al.} 2003, \apj, 593,
  1093

\bibitem[{{Macintosh} {et~al.}(2006){Macintosh}, {Graham}, {Palmer}, {Doyon},
  {Gavel}, {Larkin}, {Oppenheimer}, {Saddlemyer}, {Wallace}, {Bauman}, {Evans},
  {Erikson}, {Morzinski}, {Phillion}, {Poyneer}, {Sivaramakrishnan}, {Soummer},
  {Thibault}, \& {Veran}}]{2006SPIE.6272E..18M}
{Macintosh}, B., {Graham}, J., {Palmer}, D., {et~al.} 2006, in Presented at the
  Society of Photo-Optical Instrumentation Engineers (SPIE) Conference, Vol.
  6272, Society of Photo-Optical Instrumentation Engineers (SPIE) Conference
  Series

\bibitem[{{Makarov}(2007)}]{2007ApJS..169..105M}
{Makarov}, V.~V. 2007, \apjs, 169, 105

\bibitem[{{Marley} {et~al.}(2007){Marley}, {Fortney}, {Hubickyj},
  {Bodenheimer}, \& {Lissauer}}]{2007ApJ...655..541M}
{Marley}, M.~S., {Fortney}, J.~J., {Hubickyj}, O., {Bodenheimer}, P., \&
  {Lissauer}, J.~J. 2007, \apj, 655, 541

\bibitem[{{Mathieu} {et~al.}(2007){Mathieu}, {Baraffe}, {Simon}, {Stassun}, \&
  {White}}]{2007prpl.conf..411M}
{Mathieu}, R.~D., {Baraffe}, I., {Simon}, M., {Stassun}, K.~G., \& {White}, R.
  2007, in Protostars and Planets V, ed. B.~{Reipurth}, D.~{Jewitt}, \&
  K.~{Keil}, 411--425

\bibitem[{{McLean} {et~al.}(2003){McLean}, {McGovern}, {Burgasser},
  {Kirkpatrick}, {Prato}, \& {Kim}}]{Mclean2003}
{McLean}, I.~S., {McGovern}, M.~R., {Burgasser}, A.~J., {et~al.} 2003, \apj,
  596, 561

\bibitem[{{Mentuch} {et~al.}(2008){Mentuch}, {Brandeker}, {van Kerkwijk},
  {Jayawardhana}, \& {Hauschildt}}]{2008ApJ...689.1127M}
{Mentuch}, E., {Brandeker}, A., {van Kerkwijk}, M.~H., {Jayawardhana}, R., \&
  {Hauschildt}, P.~H. 2008, \apj, 689, 1127

\bibitem[{{Modigliani} {et~al.}(2007){Modigliani}, {Hummel}, {Abuter}, {Amico},
  {Ballester}, {Davies}, {Dumas}, {Horrobin}, {Neeser}, {Kissler-Patig},
  {Peron}, {Rehunanen}, {Schreiber}, \& {Szeifert}}]{2007astro.ph..1297M}
{Modigliani}, A., {Hummel}, W., {Abuter}, R., {et~al.} 2007, ArXiv Astrophysics
  e-prints

\bibitem[{{Mohanty} {et~al.}(2004){Mohanty}, {Basri}, {Jayawardhana}, {Allard},
  {Hauschildt}, \& {Ardila}}]{2004ApJ...609..854M}
{Mohanty}, S., {Basri}, G., {Jayawardhana}, R., {et~al.} 2004, \apj, 609, 854

\bibitem[{{Mohanty} {et~al.}(2007){Mohanty}, {Jayawardhana}, {Hu{\'e}lamo}, \&
  {Mamajek}}]{2007ApJ...657.1064M}
{Mohanty}, S., {Jayawardhana}, R., {Hu{\'e}lamo}, N., \& {Mamajek}, E. 2007,
  \apj, 657, 1064

\bibitem[{{Nordstr{\"o}m} {et~al.}(2004){Nordstr{\"o}m}, {Mayor}, {Andersen},
  {Holmberg}, {Pont}, {J{\o}rgensen}, {Olsen}, {Udry}, \&
  {Mowlavi}}]{2004A&A...418..989N}
{Nordstr{\"o}m}, B., {Mayor}, M., {Andersen}, J., {et~al.} 2004, \aap, 418, 989

\bibitem[{{Perryman} {et~al.}(1997){Perryman}, {Lindegren}, {Kovalevsky},
  {Hoeg}, {Bastian}, {Bernacca}, {Cr{\'e}z{\'e}}, {Donati}, {Grenon}, {van
  Leeuwen}, {van der Marel}, {Mignard}, {Murray}, {Le Poole}, {Schrijver},
  {Turon}, {Arenou}, {Froeschl{\'e}}, \& {Petersen}}]{1997A&A...323L..49P}
{Perryman}, M.~A.~C., {Lindegren}, L., {Kovalevsky}, J., {et~al.} 1997, \aap,
  323, L49

\bibitem[{{Saumon} \& {Marley}(2008)}]{2008ApJ...689.1327S}
{Saumon}, D. \& {Marley}, M.~S. 2008, \apj, 689, 1327

\bibitem[{{Scholz} {et~al.}(2007){Scholz}, {Coffey}, {Brandeker}, \&
  {Jayawardhana}}]{2007ApJ...662.1254S}
{Scholz}, A., {Coffey}, J., {Brandeker}, A., \& {Jayawardhana}, R. 2007, \apj,
  662, 1254

\bibitem[{{Slesnick} {et~al.}(2004){Slesnick}, {Hillenbrand}, \&
  {Carpenter}}]{2004ApJ...610.1045S}
{Slesnick}, C.~L., {Hillenbrand}, L.~A., \& {Carpenter}, J.~M. 2004, \apj, 610,
  1045

\bibitem[{{Smith} \& {Terrile}(1984)}]{1984Sci...226.1421S}
{Smith}, B.~A. \& {Terrile}, R.~J. 1984, Science, 226, 1421

\bibitem[{{Song} {et~al.}(2003){Song}, {Zuckerman}, \&
  {Bessell}}]{2003ApJ...599..342S}
{Song}, I., {Zuckerman}, B., \& {Bessell}, M.~S. 2003, \apj, 599, 342

\bibitem[{{Stelzer} \& {Neuh{\"a}user}(2000)}]{2000A&A...361..581S}
{Stelzer}, B. \& {Neuh{\"a}user}, R. 2000, \aap, 361, 581

\bibitem[{{Tinetti} \& {Beaulieu}(2009)}]{2009IAUS..253..231T}
{Tinetti}, G. \& {Beaulieu}, J.-P. 2009, in IAU Symposium, Vol. 253, IAU
  Symposium, 231--237

\bibitem[{{Tokunaga} \& {Kobayashi}(1999)}]{1999AJ....117.1010T}
{Tokunaga}, A.~T. \& {Kobayashi}, N. 1999, \aj, 117, 1010

\bibitem[{{Torres} {et~al.}(2000){Torres}, {da Silva}, {Quast}, {de la Reza},
  \& {Jilinski}}]{2000AJ....120.1410T}
{Torres}, C.~A.~O., {da Silva}, L., {Quast}, G.~R., {de la Reza}, R., \&
  {Jilinski}, E. 2000, \aj, 120, 1410

\bibitem[{{Torres} {et~al.}(2006){Torres}, {Quast}, {da Silva}, {de La Reza},
  {Melo}, \& {Sterzik}}]{2006A&A...460..695T}
{Torres}, C.~A.~O., {Quast}, G.~R., {da Silva}, L., {et~al.} 2006, \aap, 460,
  695

\bibitem[{{Torres} {et~al.}(2001){Torres}, {Quast}, {de La Reza}, {da Silva},
  \& {Melo}}]{2001ASPC..244...43T}
{Torres}, C.~A.~O., {Quast}, G.~R., {de La Reza}, R., {da Silva}, L., \&
  {Melo}, C.~H.~F. 2001, in Astronomical Society of the Pacific Conference
  Series, Vol. 244, Young Stars Near Earth: Progress and Prospects, ed.
  R.~{Jayawardhana} \& T.~{Greene}, 43--+

\bibitem[{{Torres} {et~al.}(2008){Torres}, {Quast}, {Melo}, \&
  {Sterzik}}]{2008hsf2.book..757T}
{Torres}, C.~A.~O., {Quast}, G.~R., {Melo}, C.~H.~F., \& {Sterzik}, M.~F. 2008,
  {Young Nearby Loose Associations} ({Reipurth}, B.), 757--+

\bibitem[{{van Belle} \& {von Braun}(2009)}]{2009ApJ...694.1085V}
{van Belle}, G.~T. \& {von Braun}, K. 2009, \apj, 694, 1085

\bibitem[{{Viana Almeida} {et~al.}(2009){Viana Almeida}, {Santos}, {Melo},
  {Ammler-von Eiff}, {Torres}, {Quast}, {Gameiro}, \&
  {Sterzik}}]{2009arXiv0904.1221V}
{Viana Almeida}, P., {Santos}, N.~C., {Melo}, C., {et~al.} 2009, ArXiv e-prints

\bibitem[{{Vidal-Madjar} {et~al.}(2004){Vidal-Madjar}, {D{\'e}sert},
  {Lecavelier des Etangs}, {H{\'e}brard}, {Ballester}, {Ehrenreich}, {Ferlet},
  {McConnell}, {Mayor}, \& {Parkinson}}]{2004ApJ...604L..69V}
{Vidal-Madjar}, A., {D{\'e}sert}, J.-M., {Lecavelier des Etangs}, A., {et~al.}
  2004, \apjl, 604, L69

\bibitem[{{Wallace} \& {Hinkle}(2001)}]{2001ApJ...559..424W}
{Wallace}, L. \& {Hinkle}, K. 2001, \apj, 559, 424

\bibitem[{{Zuckerman} \& {Song}(2004)}]{2004ARA&A..42..685Z}
{Zuckerman}, B. \& {Song}, I. 2004, \araa, 42, 685

\bibitem[{{Zuckerman} {et~al.}(2001){Zuckerman}, {Song}, \&
  {Webb}}]{2001ApJ...559..388Z}
{Zuckerman}, B., {Song}, I., \& {Webb}, R.~A. 2001, \apj, 559, 388

\bibitem[{{Zuckerman} \& {Webb}(2000)}]{2000ApJ...535..959Z}
{Zuckerman}, B. \& {Webb}, R.~A. 2000, \apj, 535, 959

\end{thebibliography}

\end{document}